\documentclass[aps,prl,twocolumn]{revtex4-1}
\usepackage{epsfig}
\usepackage{color}
\usepackage[normalem]{ulem}

\definecolor{red}{rgb}{1,0,0}
\definecolor{blue}{rgb}{0,0,1}
\definecolor{green}{rgb}{0,1,0}

\newcommand{\be}{\begin{equation}}
\newcommand{\ee}{\end{equation}}
\newcommand{\ba}{\begin{eqnarray}}
\newcommand{\ea}{\end{eqnarray}}

\usepackage{amsmath}
\newlength{\arrow}
\begin{document}
\title{On the density of shear transformation zones in amorphous solids}

\author{ Jie Lin${}^1$, Alaa Saade${}^1$, Edan Lerner${}^1$, Alberto Rosso${}^2$, Matthieu Wyart${}^1$}
\affiliation{${}^1$ New York University, Center for Soft Matter Research, 4 Washington Place, New York, NY, 10003, USA \\
${}^2$ Laboratoire de Physique Th{\'e}orique et Mod{\`e}les  Statistiques (UMR CNRS 8626),
    Universit\'e de Paris-Sud,
Orsay Cedex, France}

%\author{Alberto Rosso} 
%\affiliation{Laboratoire de Physique Th{\'e}orique et Mod{\`e}les  Statistiques (UMR CNRS 8626),
%    Universit\'e de Paris-Sud,
%Orsay Cedex, France}
%\author{Matthieu Wyart }
%\affiliation{New York University, Center for Soft Matter Research, 4 Washington Place, New York, NY, 10003, USA }
%
%
\date{\today}

\begin{abstract}
We study the stability of  amorphous solids,  focussing on the distribution $P(x)$ of the local stress increase $x$ that would lead to an instability.
We argue that this distribution is singular $P(x)\sim x^\theta$,
where the exponent $\theta$ is non-zero if the elastic interaction between rearranging regions is non-monotonic, and increases 
with the interaction range. For a class of finite dimensional models we show that stability implies a lower bound on $\theta$, which  is found to lie near saturation. For quadrupolar interactions these models yield $\theta\approx {0.6}$ for $d=2$ and $\theta\approx 0.4$ in $d=3$ where $d$ is the spatial dimension, accurately capturing previously unresolved observations in atomistic models,  both in quasi-static flow and after a fast quench. 

%These results are independent of the choice of dynamical rules chosen, which however affects  rheological properties, in particular the Herschel-Bulkley relation.  Our findings suggests an explanation for puzzling differences between  the depinning transition and the still poorly understood yielding transition, and  supports that popular models of plastic flow accurately capture non-trivial features of the glass transition, where the role of long-range elastic interactions is generally not considered important. 
\end{abstract}


\maketitle

%One of the hallmark of amorphous materials  is the abundance of soft excitations associated with   local rearrangement  of particles. 
%Such excitations were detected in the 70's in low-temperature measurements of the specific heat in glasses \cite{zeller}, which support that the dominant low-energy excitations are not phonons as is the case in crystals, but rather two-level systems where a group of atoms can tunnel between two distinct configurations \cite{phillips_book}. 

Dislocations play a key role in controlling plastic flow  in crystalline solids. In contrast, the notion of defects  is ill-defined in amorphous materials.  However plasticity under shear in these materials occurs via  events that are  also well localized in space \cite{argon, maloney,Picard2004}. 
%One such event  results in a quadrupolar perturbation  of stress  in the far field \cite{maloney,Picard2004}. 
There are thus preferential locations where plastic rearrangements are  likely to occur, which have been coined shear transformation zones (STZ) ~\cite{argon}%\sout{zones (STZ)}
, and are central to various proposed descriptions of plasticity \cite{langer98}.  The microscopic nature of these objects is however elusive, and their concentration is thus hard to measure directly \cite{Rodney,Manning}.  
%Interestingly, a similar state of affairs occurs near the glass transition. It is found that the thermally-activated structural relaxation is heterogeneous and occurs in some favored locations or soft spots \cite{harrowell2} where soft vibrational modes are abundant \cite{brito07b,manning2}.  %It is presently unclear if there is a connection between STZ and soft spots.
Recently it has been shown \cite{Wyart2012,lernersoft}  in the particular case of packings of hard particles  that the density $P(x)$ of  excitations  that rearrange if a local additional stress $x$ is  applied  is singular:  $P(x)\sim x^\theta$,
where the exponent $\theta$ is tuned such that small perturbations can have dramatic effects. This result raises the question of how stable  generic amorphous solids are, and how this stability is reflected  in the distribution of excitations $P(x)$. There is indirect evidence that $P(x)$ is indeed singular even when smooth interaction potentials are considered \cite{itamar}: both following a quench and during steady flow, the increment where stress can be increased without plastic events in a system of $N$ particles  scales as $N^{-\xi}$ where $\xi<1$. Assuming that the rearranging regions are independent  would imply that $\theta>0$, as we shall recall below. However, the hypothesis of independence is inconsistent with observations at the yield stress~\cite{itamar}, raising doubts on the inference of $\theta$. Most importantly, what controls this singularity in the density of excitations is not known. 

In this Letter we argue that the exponent $\theta$ is governed by the interaction between rearranging regions.  $\theta$ can be non-zero if the interaction is non-monotonic, i.e. is either stabilizing or destabilizing depending on the location, and increases 
with increasing interaction range. We extend a previous mean-field, on-lattice model of plasticity \cite{lequeux} to the case of  power-law interactions, and show that stability implies a lower bound on~$\theta$, which  is found to lie near saturation.  %These results are independent of the exact choice of the dynamics, which however affects the relationshipw betwe
When more realistic quadrupolar interactions are considered, which are known to characterize the far field effect of a plastic event \cite{maloney,Picard2004}, the model yields $\theta\approx {0.6}$ for $d=2$ and $\theta\approx 0.4$ in $d=3$, and reproduces at a surprising level of accuracy the system size dependence of the strain interval between plastic events observed in atomistic models \cite{itamar}, {\it both} in flow and after a fast quench.
Our findings suggest an explanation for puzzling differences between  the depinning transition where an elastic manifold is driven in a random environment, and the yielding transition. They also  support  that popular models of plastic flow capture some aspects of the dynamics going on at the glass transition. 

%The small force distributions in amorphous material determine the mechanical properties when the material is about to yield\cite{Wyart2012} \cite{Lerner2012}. It's well believed that atoms in different positions interact through anisotropic propagator\cite{Maloney2009} \cite{Maloney2004}, and the propagator expression in two dimension is quadrupole and decay with $1/r^2$. Unlike depinning, the local deformation event always destablizes others sites to yield as well, the effect of local deformation could stablize or destablize other atoms at the same time in plasticity. We show that this qualitive difference result in the non-trival small stresses distributions with a power law form, instead finite distributions in depinning cases \cite{Fisher}.

We consider cellular automaton models which are known to capture the critical behavior of the depinning transition \cite{Aragon2012} and have been used to study plastic flow \cite{lequeux,Baret2002,Zaiser2006,ajdari2009,12MBB}.
 The sample is decomposed into sites labeled by $i$, each carrying  a scalar shear stress $\sigma_i$. A site represents a few particles.  The applied shear stress is $\sigma=\frac{1}{N} \sum_{i} \sigma_i$. 
For each site we define a local threshold $\sigma_i^{th}$, which for simplicity is taken to be unity. If   $\sigma_i>\sigma_i^{th}$, the site is mechanically unstable.  When unstable, a site has  a probability per unit time $1/\tau_c$ to rearrange plastically, in which case the local stress is set to zero.  Physically, $\tau_c$ is the characteristic  time to relax toward a new local minimum of energy. 
Such a plastic event  affects the stress in the rest of the system, with a delay associated to elastic propagation. We shall neglect that this delay depends on the distance from the plastic event, and choose it to be exponentially distributed with mean $\tau_r$.
% We shall make the simplification that  both the characteristic time and the delay time are exponentially distributed of mean  $\tau_c$  and $\tau_r$ respectively, and that $\textcolor{green}{\tau_c},\tau_r$ and $\sigma_i^{th}=1$ are position independent. 
A plastic event  leads to the following change in the local stresses:
% at a rate $\tau_c$. $\tau_c$ thus characterizes the speed at which a local irreversible instability occurs. An instability that occurs at some time $t$ will affect  the stress in the entire system, with some delay. We shall make the simplification that this  delay $\tilde \tau_r$ is exponentially distributed and of mean $\tau_r$, and is independent of position.  During the time interval $[t,t+\tilde\tau_r]$, . At a time $t+\tilde \tau_r$,   local stresses  change as follows:
\ba
\sigma_i&\rightarrow& 0\\
\sigma_{j }&\rightarrow& \sigma_{j} + {\cal G}({\vec r}_i-{\vec r}_j).
\ea
 %We define  
 %yield stress $\sigma_{y}$ of amplitude unity $\sigma_y\equiv 1$, such that at any instant of time if a site $i$ satisfies $\sigma_i>\sigma_{y}$, it becomes  mechanically {\em unstable}. 
 %If the site $i$ remains unstable for a time interval drawn from an exponential distribution of mean $\tau_c$,  the site $i$ is said to be in a {\em collapsed state} which a
 In simple depinning models,  ${\cal G}$ is strictly positive and the interaction is said to be monotonic.  We focus here on non-monotonic interactions for which the sign of ${\cal G}$ varies. In this case if $\tau_c>0$, unstable sites can be re-stabilized by other plastic events.
 Such models predict the existence of a yield stress, and as we shall see are consistent with the Herschel-Bulkley relation:
 \be
 \label{1} 
 \dot\gamma\sim (\sigma-\sigma_c)^\beta
 \ee
where the strain rate $\dot{\gamma}$ is defined as the number of collapses per unit time.
In numerical simulations   finite-size fluctuations  can stop the dynamics even if $\sigma>\sigma_c$. When this happens we give small random kicks to every site until a new site becomes unstable. This method  enables us to reach the steady state when $\sigma>\sigma_c$ and  to study avalanche dynamics at $\sigma_c$.
 %which is in general a much less understood problem
 %~\cite{Fisher}. \textcolor{blue}{WE WOULD LIKE TO KEEP THE DESCRIPTION MINIMALISTIC BUT COMPLETE. FEEL FREE TO ADD/MODIFY AS NEEDED.} %could be infinite long range fluctuation, power law decay, and quadrupole interaction.

We shall denote the  distance to the yield stress of the site $i$ by $x_i \equiv 1-\sigma_i$. Our goal is to understand how the distribution $P(x)$  depends on the interaction  ${\cal G}$.  We introduce the decomposition $P(x)=P_1(x)+P_2(x)$ where $P_2(x)$ corresponds to collapsed sites, and $P_1(x)$ to the other sites.
%We define the strain rate $\dot{\gamma}$ as the rate at which plastic events occur per particle, which follows:
%\begin{equation}
%\label{1}
%\dot{\gamma}=\frac{\int_{-\infty}^{\infty} P_2(x) dx}{\tau_r}=\frac{\int_{-\infty}^{0}P_1(x) dx}{\tau_c}
%\end{equation}
%The last equality simply states that in a stationary state, the flux of collapsing site is equal to the flux of sites that become stabilized again. 
%is because the rate of $s_i=1 \rightarrow -1 $ must be equal to the rate of $s_i=-1\rightarrow 0$ since the fraction of inactive sites is constant in steady states and $\int_{-\infty}^{\infty}P_2(x) dx=1-\int_{-\infty}^{\infty}P_1(x)dx$. Note these definitions of strain rate are universal and independent on the choice of propagator.

\begin{figure} [ht]
{\includegraphics[width=.99\linewidth]{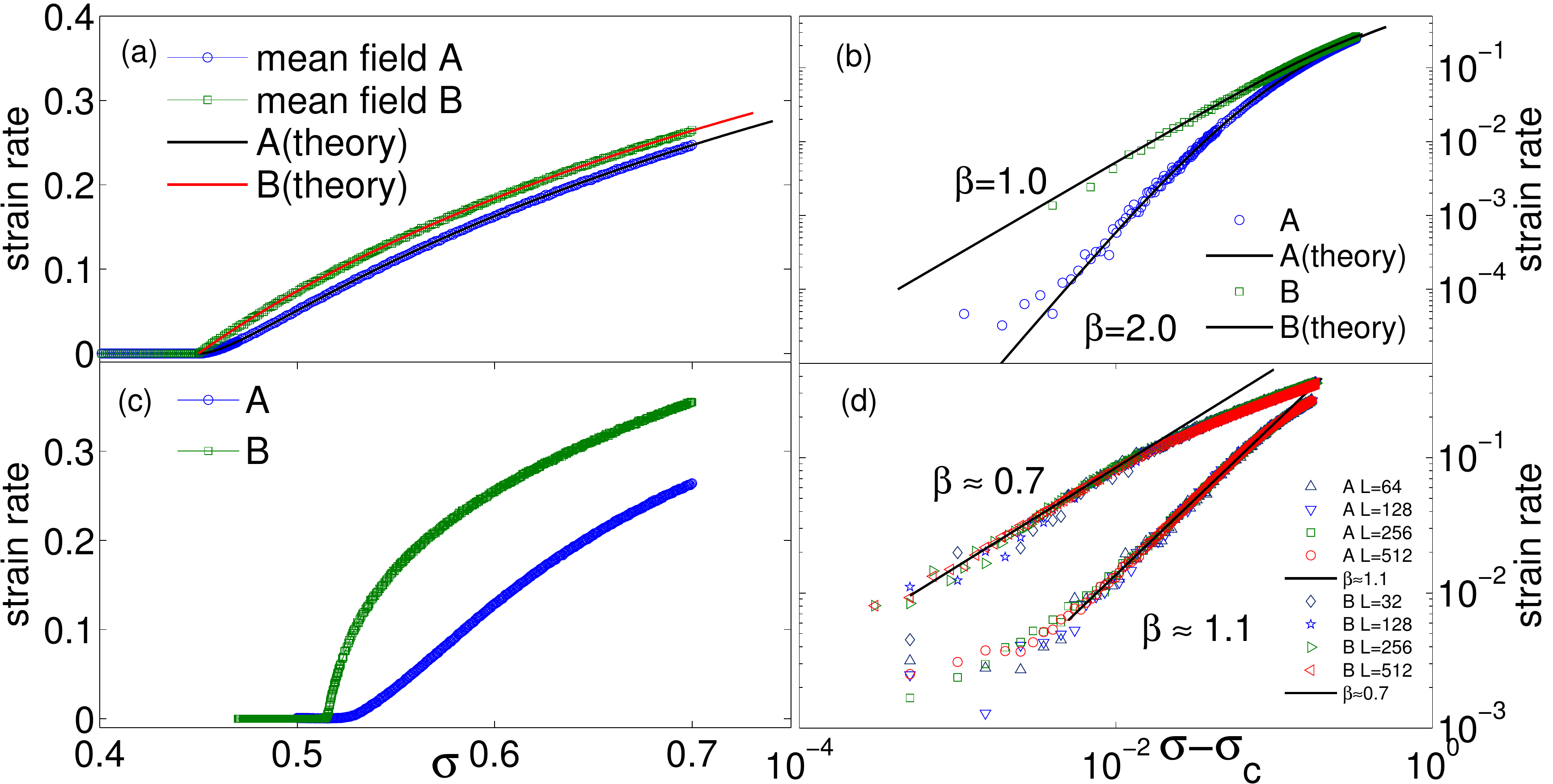}}\\
\caption{\small{(color online). ${\dot{\gamma}}$ {\it v.s.}~$\sigma$ for Models A and B (see text for details). Top panels: mean field interactions with $D= 1/6$ on (a) linear and (b) log-log scales. Bottom panels: quadrupole interactions for $d=2$ on (c) linear and (d) log-log scales. The fits give $\beta \approx 1.1$, $\sigma_c \approx 0.52$ for model A and $\beta \approx 0.7$, $\sigma_c \approx 0.515$ for model B.}}
\label{meanss}
\end{figure}

We first consider a solvable mean-field model where  it is assumed that a plastic event leads to random kicks of stress that do not depend on position: 
\begin{equation}
\label{2}
{\cal G}_{mean}({\vec r}_i-{\vec r}_j)=\frac{\eta_j }{\sqrt{N}} + \frac{\tilde \eta}{N}
\end{equation}
where the $\eta_j$ are independent variables in space, and are not correlated from one plastic event to the next. They are randomly distributed in $[-\eta_0, \eta_0]$, and $\eta_0$ does not depend on $N$ to ensure the existence of a thermodynamic limit. $\tilde \eta$ is chosen at each collapse event to ensure that the total stress is conserved, and thus depends on the random variables $\eta_j$ of that event. This model  is a slight variation  of that of Hebraud and Lequeux \cite{lequeux}, and we shall briefly recall how to solve it. 
%This study will enable us to discuss in a simple case the physics what governs the distribution of nearly unstable regions. We shall also see that, in contrast to the case of monotonic interactions, a priori innocent choices of dynamical rules can affect qualitatively relationship between strain and stress, but not the statistics of avalanches nor the distribution $P(x)$.  

In the thermodynamic limit Eq.(\ref{2}) implies a Fokker-Planck equation for active  sites:
\be
\frac{\partial P_1(x)}{\partial t}=\\
\dot\gamma\Big(D\frac{\partial^2P_1(x)}{\partial x^2}+\lambda \frac{\partial P_1(x)}{\partial x}+\delta(x-1)\Big)-\theta(-x)\frac{P_1(x)}{\tau_c} \label{FP1}
\ee
and an equation of similar form for $P_2(x)$ (see Appendix). $D\equiv\frac{\eta_0^2}{6}$ represents the diffusion constant of the local stress, coming from the random kicks of other collapsing sites.  $\lambda$ is a Lagrange parameter that constrains the average stress, and is chosen such that $\int xP(x)dx=1-\sigma$. It results from the second term on the RHS of Eq.(\ref{2}). The $\delta$ function term corresponds to the flux of reinserted sites at $\sigma=0$, equivalent to $x=1$. The last term in Eq.(\ref{FP1}) corresponds to the flux of unstable sites that collapse,  and $\theta(x)$ is the Heaviside function. Eq.(\ref{FP1}), together with a similar equation for $P_2(x)$, are closed.
%Note if there is no fluctuation, the model is similar to the previous mean field proposed by Dahmen, et al \cite{Dahmen2009} \cite{Dahmen1998}.
%We find that no stationary solutions with $\dot\gamma$ \textcolor{blue}{$>0$} exist for $\lambda<1$. At $\lambda=1$, the critical distribution $P_c(x)$ is independent of $\tau_c$ and $\tau_r$ and follows:
%\begin{align}
%P_c(x) & = 0  & x<0  \notag\\
%P_c(x) & = 1-\exp{(-x/D)}   & 0<x<1  \label{3}\\
%P_c(x) & = (\exp{(1/D)}-1) \exp{(-x/D)}   & x>1\notag
%\end{align}
%Taking the mean one gets $x_c=D+\frac{1}{2}$, implying for the yield stress $\sigma_c=\frac{1}{2}-D$.  
%Eq.(\ref{3}) implies that $P_c(x)\sim x$, \sout{or} \textcolor{blue}{i.e.}~$\theta=1$ in mean-field. 
We find (see Appendix) that no stationary solution with $\dot{\gamma}>0$ exists for $\sigma<\sigma_c=1/2-D$. The critical distribution ($\sigma=\sigma_c\implies\dot{\gamma}=0$) is independent of $\tau_c$ and $\tau_r$, and satisfies $P(x)\sim x$ for small $x$.
This linear density of nearly unstable regions has a simple explanation:
the stress in each site follows a diffusion equation. In the limit of small strain rate the instability threshold at $x=0$ becomes an absorbing boundary condition. 
In contrast, for monotonic problems such as pinned elastic interfaces,  the distance to a local instability $x$ is always decreasing in time and $P(x)$ thus cannot vanish at $x=0$.
\begin{figure}[!hbt]
  {\includegraphics[width=1.0\linewidth]{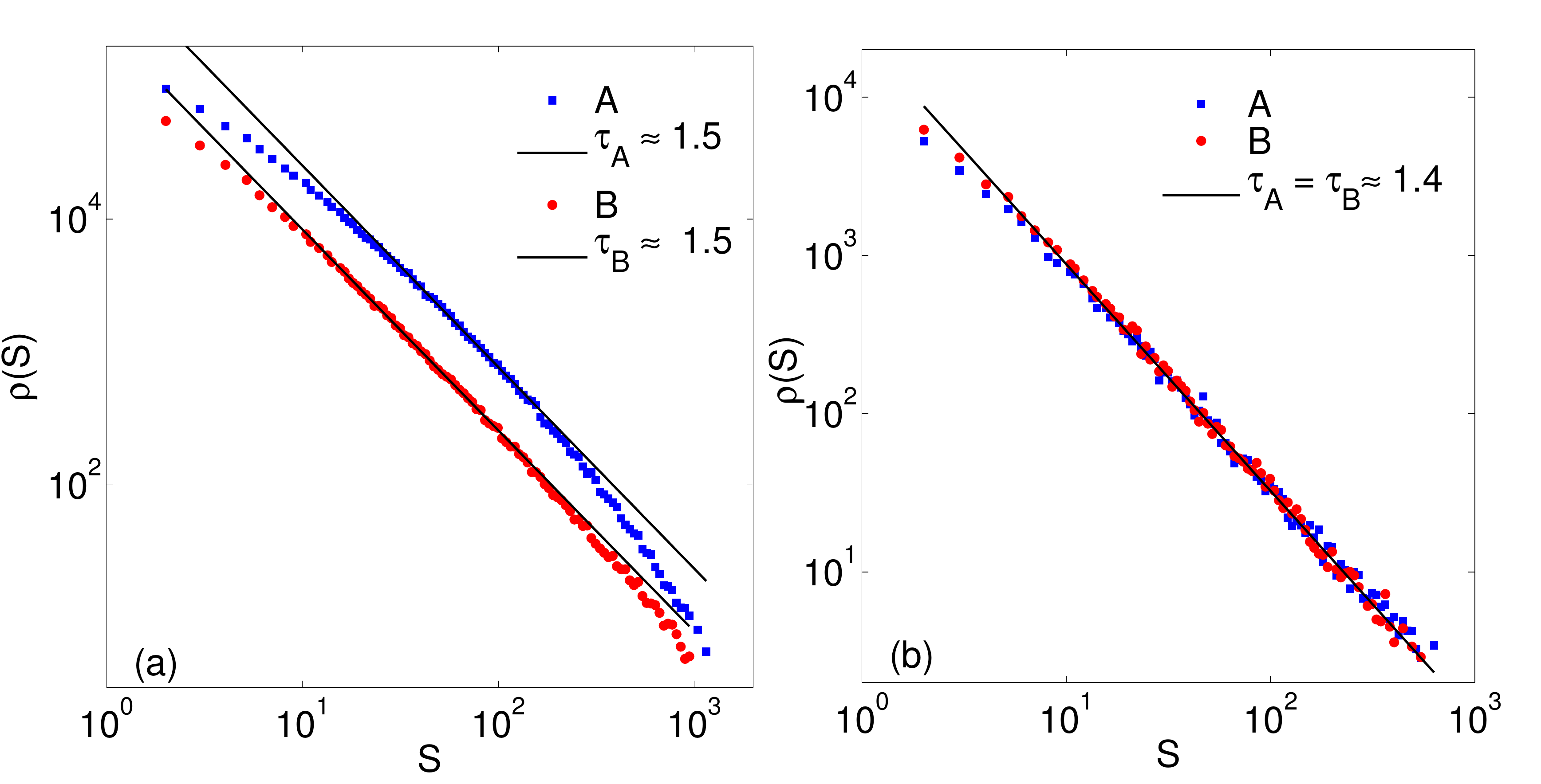}}
\caption{\small{Distribution of avalanches size for  (a) mean-field interaction and (b) quadrupole interactions.}}\label{ava}
\end{figure}

Solving this mean field model at finite strain rate, we find at first order in $\dot \gamma$:
\begin{equation}
\label{scalingMF}
\begin{split}
&\sigma -\sigma_c =\\
&(\frac{1}{2}+D)\sqrt{D \tau_c} \dot{\gamma}^{1/2}+[	\tau_r(\frac{1}{2}+2D)+\frac{\tau_c}{4}+(1-D)D\tau_c]\dot{\gamma}
\end{split}
\end{equation}
We thus see that if $\tau_c\neq0$, $\sigma-\sigma_c\sim \dot\gamma^{1/2}$, corresponding to  $\beta=2$ in Eq.(\ref{1}), also obtained in the model of \cite{lequeux}. However if 
$\tau_c=0$, $\sigma-\sigma_c\sim \dot\gamma$, corresponding to $\beta=1$. These results are tested numerically for Model A ($\tau_r=0,\tau_c=1$) and Model B ($\tau_r=1,\tau_c=0$) in Fig.(\ref{meanss}). Despite the notable difference between $A$ and $B$, $P_c(x)$ is independent of the choice of dynamics. This supports that the class of models we consider should yield correct values for $\theta$, but indicates that our assumption that the delay $\tau_r$ is independent of position may yield incorrect Herschel-Bulkley exponents. Interestingly, we find at $\sigma_c$ that the avalanches distribution $\rho(S)$, where $S$ is the number of plastic events,  
%(avalanches are initiated by giving little random kicks to trigger a first plastic event) 
follows $\rho(S)\sim 1/S^{\tau}$ with $\tau\approx3/2$  independently of the choice of dynamics, as shown in Fig.(\ref{ava}). This result is consistent with mean-field depinning \cite{fisher} and the ABBM model \cite{ABBM}.

%
%We also look at the avalanche distribution close to the critical stress by inducing random fluctuation to absorbing system with no unstable sites, where the avalanche size is defined as the number of sites collapsing during an avalanche. In both dynamics, we find power law distribution of avalanches $P(S) \sim S^{-\tau}$, with $\tau_A\approx 1.4$ and $\tau_B\approx 1.5$,  close to the mean field value proposed by Dahmen, et al, see Fig.(\ref{ava}a).
%
%
%We see that even the most simple fluctuation mean field model gives us a non-zero exponent $\theta$. We argue it's because the propagator we use has stabilization and destablization effects at the same time, the stress of site which is close to the local yield stress somehow diffuse near the yield stress rather than keep going and become closer and closer to the yield stress. The diffusion requires the stress distribution must be zero at the local yield stress to be continuous, and asymptotically we must obtain power law decay of stress distribution near the local threshold stress.

\begin{figure}[!hbt]
{\includegraphics[width=1.0\linewidth]{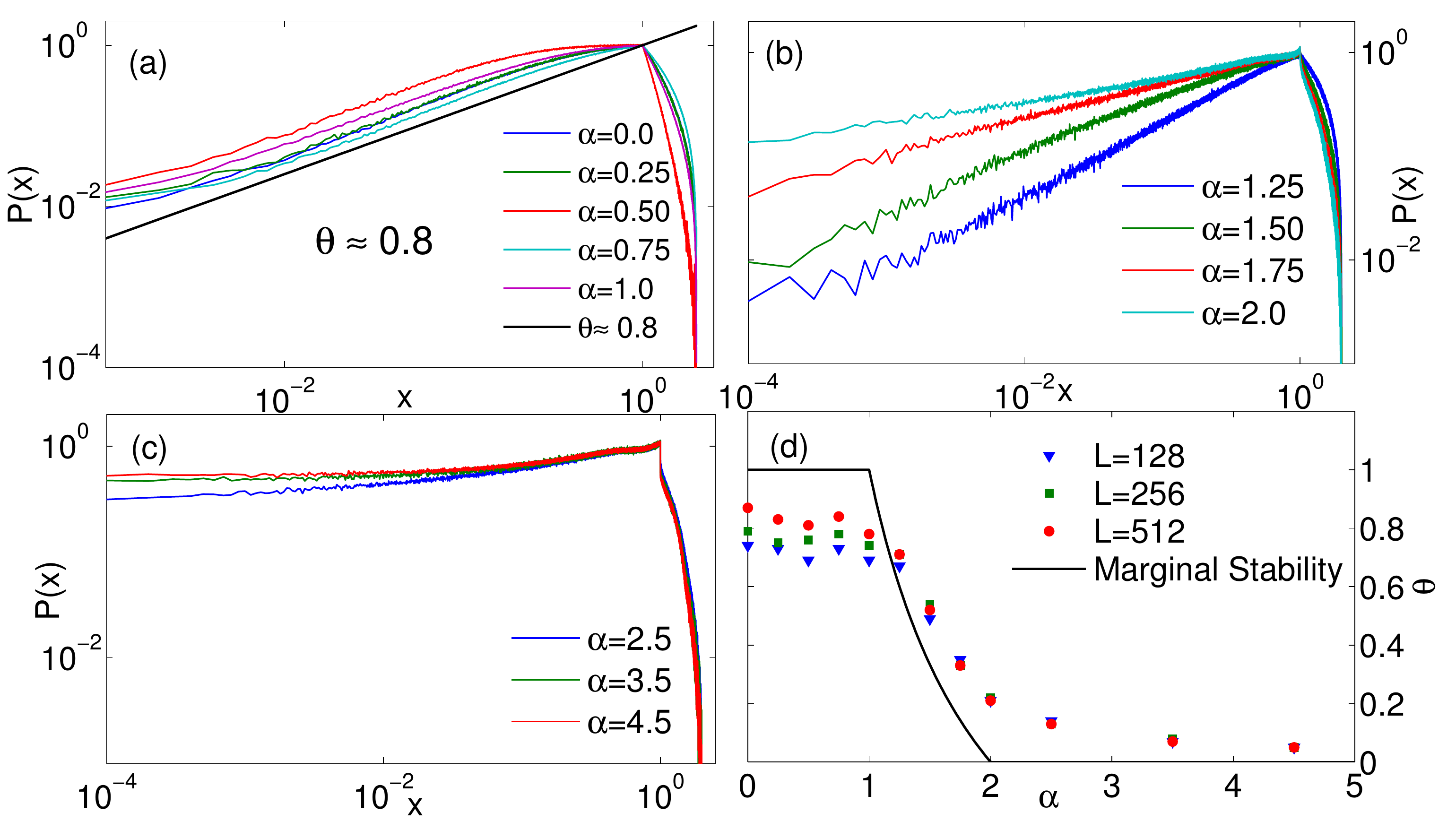}}
\caption{\small{(color online). $P(x)$ {\it v.s.}~$x$ for a $d=2$ system with power-law interaction 
of Eq.(\ref{5}): (a) When $\alpha \leq 1$, $\theta \ge 0.8$  (b)  When $1<\alpha \leq 2$,  $\theta$ decreases as $\alpha$ increases. (c) When $\alpha>2$,  $\theta \approx 0$. (d) $\theta$ {\em vs} $\alpha$ obtained in our simulation and compared with the theoretical lower bound of Eq.(\ref{7}).
%of exponent $\alpha$ with (a) $\alpha \leq 1$, where $\theta \approx 0.8$  (b)  $1<\alpha \leq 2$, where $\theta$ decreases slowly as $\alpha$ increases. (c) $\alpha>2$, showing that $\theta$ is above but close to zero. (d) $\theta$ as a function of $\alpha$ is compared with the theoretical lower bound.
}}\label{powertheta}
\end{figure}

To study finite dimensional effects, we now consider interactions decaying with distance, of the form:
\begin{equation}
\label{5}
{\cal G}_{mean}({\vec r}_i-{\vec r}_j)=\frac{\eta_j }{r_{ij}^\alpha} +  \eta_1
\end{equation}
where $\eta_j\in[-\eta, \eta]$ is a random variable  uniformly distributed, and $\eta_1$ is again a global shift to keep the average stress constant.
% Note that the fluctuation mean field corresponds to the $\alpha\rightarrow 0$ limit. To get the stress distribution at the critical stress directly, we neglect the second term in (\ref{power}), and make use of the competition of stress decrement from reinsertion and stress increment from inducing avalanche to make the system evolving into the critical states automatically \cite{Talamali2011}.
%
%
To ensure that the diffusion constant stemming from plastic events has a
  thermodynamic limit, the coefficient $\eta$ must be such that $\int_1^L \frac{\eta^2}{r^{2\alpha}} d^{d}r\sim1$ where $d$ is the spatial dimension and $L$ the linear system size. 
We get $\eta \sim L^{\alpha-d/2}$ for   $\alpha<d/2$,   $\eta \sim 1/\sqrt{\ln L}$ for $\alpha=d/2$ and    $\eta \sim 1$  for      $\alpha>d/2$.

Computing $\theta$ is now a much harder problem. However we now show that stability implies a bound on this exponent. Let us denote by $m$ the average number of plastic events that are triggered if one  single event at the origin takes place. We assume that the distribution $P(x)$ satisfies $P(x)\sim x^\theta$. A site at a distance $r$ experiences a  kick which contains the term  $\eta_1$ of order $1/N$ that stems from stress conservation and a term $\eta/r^\alpha$, which is destabilizing or stabilizing with probability $1/2$.  The term $\eta_1 \sim 1/N$ will destabilize the site with a probability $p_1\sim P(x<1/N)\sim N^{-(1+\theta)}$, so that overall in the entire system this will trigger of the order of $N^{-\theta}$ events,  which is negligible as long as $\theta>0$.  The probability $p_2(r)$ that the term  $\eta/r^\alpha$  destabilizes the site is of order $p_2(r)=P(x\leq \eta/r^\alpha)/2\sim\eta^{\theta+1}/r^{\alpha(\theta+1)}$. Integrating over all sites we get:

% a constant part of oder $1/N$ that stems from stress conservation -the last term of Eq.(\ref{5})- and a random one of order $\eta/r^\alpha$, which is destabilizing or stabilizing with probability $1/2$.  The constant kick will destabilize the site with a probability $p_1\sim P(x<1/N)\sim N^{-(1+\theta)}$, so that overall in the entire system this will trigger of the order of $N^{-\theta}$ events,  which is negligible as long as $\theta>0$.  The probability $p_2(r)$ that the random part of the kick destabilizes the site is of order $p_2(r)=P(x\leq \eta/r^\alpha)/2\sim\eta^{\theta+1}/r^{\alpha(\theta+1)}$. Integrating over all sites we get:
\begin{equation}
m\sim \eta^{\theta+1} \int_1^L \frac{dr}{r^{\alpha(\theta+1)+1-d}}\sim \eta^{\theta+1} L^{d-\alpha(\theta+1)}\sim L^\nu\label{a4}
\end{equation}
where the exponent $\nu$ can be computed from the dependence of $\eta$ with $L$. Stability toward run-away avalanches requires $\nu\leq0$, which finally leads to:
\ba
\label{7}
\theta&\geq& 1\ \ \ \ \ \ \ \ \ \hbox{     if} \ d\geq 2\alpha \nonumber \\
\theta&\geq&\frac{d}{\alpha}-1\  \ \  \hbox{    if} \ d/2\leq\alpha\leq d\\
\theta&\geq& 0 \ \ \ \ \ \ \ \ \ \ \hbox{     if } \ \alpha>d \nonumber
\ea
This prediction is tested in Fig.(\ref{powertheta}), where $P(x)$  is shown for various interaction range $\alpha$ in two dimensions, from which the exponent $\theta$ is extracted.  Fig.(\ref{powertheta}.d) shows the comparison between the measurement and the stability bound. 
 For  $\alpha<1$ the bound appears slightly violated, but less and less so for larger system sizes, supporting that $\theta=1$, as we predicted exactly for $\alpha=0$. For  larger $\alpha$, $\theta$ is found to lie close to the bound but systematically above, although our data cannot rule out saturation. 
%The bound appears slight violated for $\alpha<1$, but less and less so for larger system sizes, supporting that $\theta=1$ the bound is saturated -as we predicted exactly for $\alpha=0$. For  larger $\alpha$, $\theta$ is found to lie close to the bound but systematically above, although our data cannot rule out saturation. 
Our analysis thus implies that the range of interaction is a key determinant of $P(x)$ and $\theta$, and that for this class of models,  systems lie close to marginal stability.  

%Note the first term is from the power law part and second term is from the global shift. Combining (\ref{eta}-\ref{a4})  and $d=2$, we get
%\begin{align}
%\theta&=1  &0<\alpha<1 \\
%\theta&\geq \frac{2}{\alpha}-1  &1<\alpha<2 \\
%\theta&>=0  &\alpha>2 
%\end{align} 

%Note that the weakest site has a typical stress $1-\sigma_m\sim N^{1/(\theta+1)}$. From numerical simulations, we find $\theta \approx 0.85$ when $\alpha \leq 1$ for $L=512$, and there is tendency to get theoretical value $\theta=1$ in the thermodynamic limit. When $\alpha>1$  $\theta$ decays to zero as $\alpha$ increase and the curve is above the marginal stability prediction line as expected. When $\alpha>2$, one could think $\theta \approx 0$, see Fig.\ref{poalberto_rossowertheta}.
% \begin{figure} [!hbt]
% {\includegraphics[width=1.0\linewidth]{figure/stressstrainratecombinenew}}
% \caption{\small{Strain rate {\it v.s.} stress for the quadrupolar interaction in two dimensions for dynamics A, where $\sigma_c=0.527$ and $\beta_A\approx 1.1$, and for dynamics B, where  $\sigma_c=0.517$ and $\beta_B\approx0.7$. }}\label{quadss}
% \end{figure}

Finally we consider the more realistic case where the interaction is not random, but quadrupolar in the far field \cite{Picard2004}. Our model then belongs to the class of elastoplastic models \cite{ajdari2009,Talamali2011,Zaiser2006,12MBB}. In two dimensions for a simple shear along the $x$ axis  one has for  an infinite system ${\cal G}(r)=\frac{\cos{4\phi}}{r^2}$, where $\phi$ is the angle made with the $x$ axis. Periodic boundary conditions can be implemented using the Fourier representation, ${\cal G}(k_x,k_y)=k_x^2 k_y^2/k^4$, and  the discrete wave vectors, $k_x=2 \pi n_x/L$,$k_y=2 \pi n_y/L$.
This interaction can be computed also in $d=3$ where it decays as ${\cal G}(r)\sim 1/r^3$. 

\begin{figure}[!hbt]
\centering\includegraphics[width=1.0\linewidth]{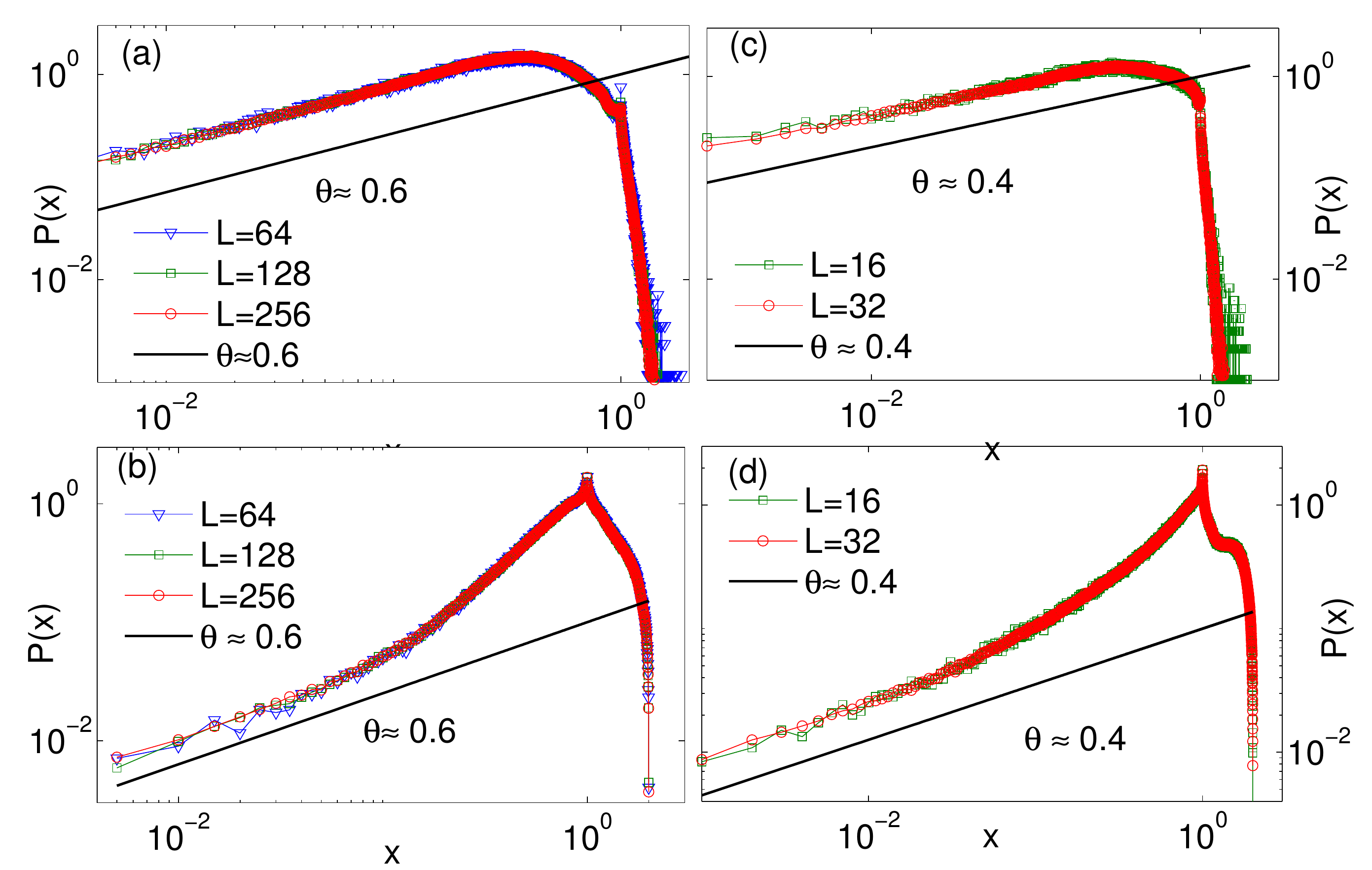}
\caption{\small{(color online). $P(x)$ {\it v.s.}~$x$ (a) for $d=2$ at the critical stress, (b) for $d=2$ after a fast quench (c) for $d=3$ at the critical stress and (d) for $d=3$ after a fast quench. Results are shown for dynamics A, and are identical for dynamics B.}}\label{quadtheta}
\end{figure}

%We take bi-periodic boundary condition, so the wave vector is constrained at discrete values, $k_x=\frac{2 \pi n_x}{L}$,$k_y=\frac{2 \pi n_y}{L}$. We point out that $g(k_x=0,ky\neq0)=0$ and $g(k_x \neq 0,ky=0)=0$  which means the sum of stress increment on each row and column is exactly zero! To be realistic and close to the thermodynamic limit, we impose the sum of stress over each column and row are exactly the same as $\langle \sigma \rangle$ for the initial condition.

To our knowledge the exponent $\beta$ of Eq.(\ref{1}) has not been computed for such models.  Our results are shown in Fig.(\ref{meanss}) bottom for dynamics A and B, and are well-captured by the Herschel-Bulkley law with $\beta_A\approx 1.1$ and $\beta_B \approx  0.7$.  We also compute the avalanche exponent and find $\tau \approx 1.42$ in both dynamics, again close to the mean field value $1.5$ in agreement with \cite{Zaiser2006} but somewhat larger than \cite{Talamali2011}, as shown in Fig.(\ref{ava}).

% \cite{Sun2010}  %\cite{Salman2011}\cite{Weiss2001}, see .  
\begin{figure}[!hbt]
\centering\includegraphics[width=1.0\linewidth]{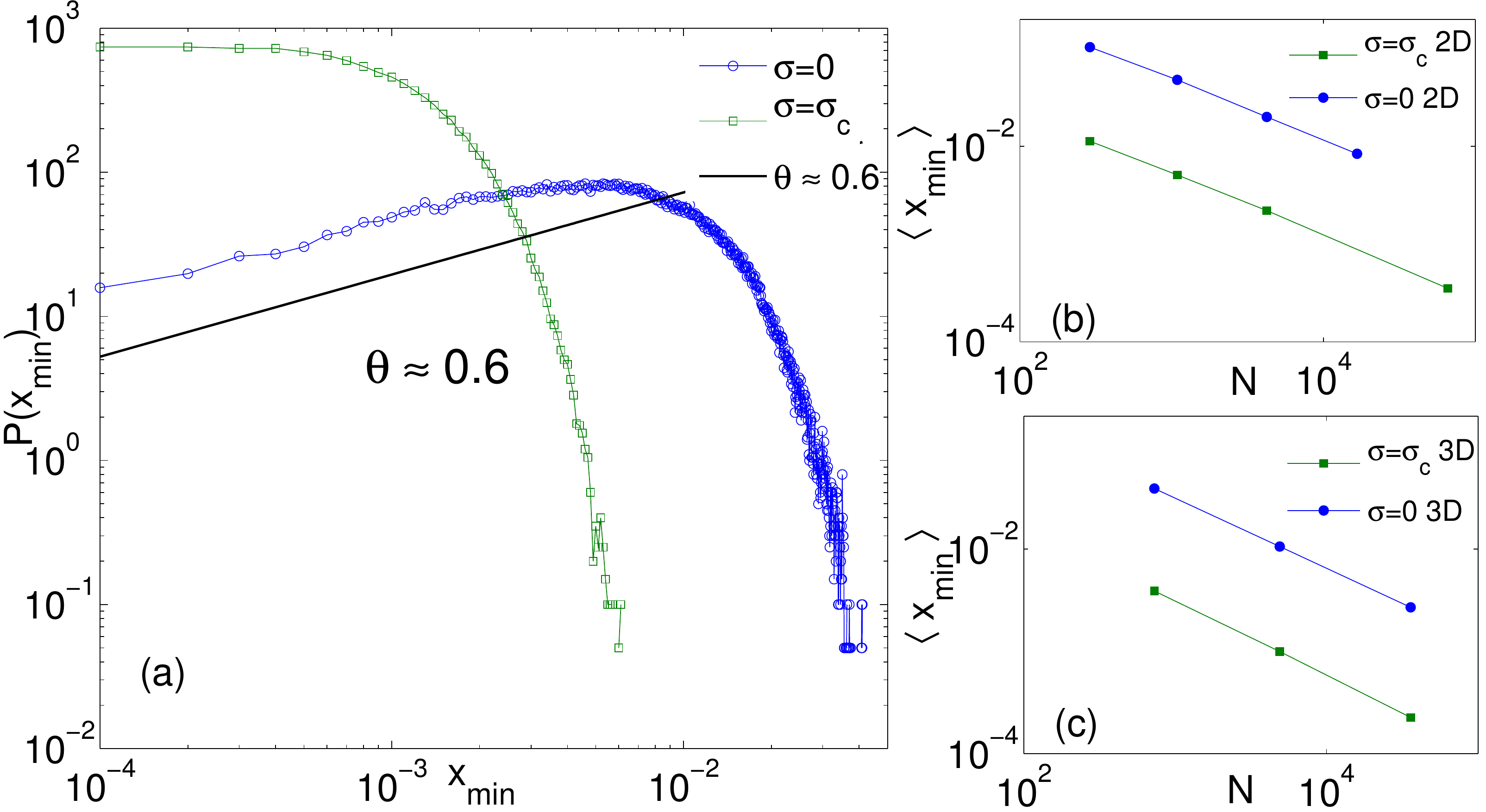}
\caption{(color online). Left: distribution of the most unstable site $P(x_{\rm min})$ at $\sigma_c$ (green square) and  after a rapid quench, at $\sigma=0$,  (blue circles)  for $d=2$. Right: Evolution of $\langle x_{\rm min}\rangle$ {\it v.s.} $ N$ indicating a power law regime $\langle x_{\rm min}\rangle \sim N^{-\xi}$ with $\xi\approx0.63$ in $d=2$ and $\xi\approx0.71$ in $d=3$ both for $\sigma_c$ and the rapid quench.  }\label{xmin}
\end{figure}

One central result concerns the  density of excitations $P(x)$. We measure this quantity in two situations: (i) at the yield stress $\sigma_c$, in the steady state  (ii)  at $\sigma=0$ after a "quench" which mimics the behavior that would occur if the temperature was suddenly set to zero in a liquid. In the latter case the {\em initial} local stress, $\sigma_i$,  are drawn from a Gaussian symmetric distribution  $Q_0(\sigma)$, so that $\sigma=0$.  The system is however unstable because many sites with $|\sigma_i|>1$ can collapse  and trigger other rearrangements. This dynamics stops when $|\sigma_i|<1$ on all sites. Our results are shown in Fig.(\ref{quadtheta}). We find that $\theta\approx 0.6$ in two dimensions and $\theta\!\approx\! 0.4$ in three dimensions, both at $\sigma_c$ and after the quench at $\sigma=0$.
 %.  of zero mean ($\sigma=0$), but with and variance broader than the 
%The quench consists in preparing the system with an initial  distribution of local stress $Q(\sigma)$, of zero mean but  of variance broader than the yield stress. The dynamics that follows is governed by our model, for which any local stress satisfying  $|\sigma_i|>1$ can collapse, and trigger other rearrangements. , 

Can such a simple model, that neglects in particular the tensorial nature of stress, % and assumes anisotropic interaction, 
capture essential aspects of the glassy dynamics that occurs during an isotropic quench? To test this hypothesis we compare our results with the atomistic simulations of \cite{itamar}.  $P(x)$ is not available directly, but the statistics of $x_{\rm min}$ can be obtained accurately by considering the minimal increment of strain, or stress, required to generate a plastic event. It is found that $\langle x_{\rm min}\rangle\sim N^{-\xi}$ with $\xi\!\approx\! 0.62$ after a quench, and $\xi\!\approx\! 2/3$ at the yield stress, with no clear dependence on the dimension. Our measurement of the exponent $\xi$ are shown in Fig.(\ref{xmin}), and are remarkably similar to these observations, as we find  $\xi\!\approx\!0.63$ in $d=2$ and $\xi\approx0.71$ in $d=3$ both  after a quench at $\sigma=0$ or in the steady state at $\sigma_c$. The exponent $\xi$ can be related to $\theta$ if one assumes  the independence of the variables $x_i$. %In that case $\langle x_{\rm min}\rangle$ can be easily estimated in the large $N$ limit: 
\begin{equation}
 \int_0^{\langle x_{\rm min}\rangle} \, d x\, P(x) \sim \frac{1}{N}  \;  \to \; \langle x_{\rm min}\rangle \sim N^{-\frac{1}{\theta+1}}
\end{equation}
leading to $\theta=1/\xi-1$, a relationship satisfied in our data. The independence assumption also implies a specific form for the distribution of the most unstable site at small argument $P(x_{\rm min})\!\sim\!x_{\rm min}^\theta$, which is indeed observed in~\cite{itamar} for quenched systems, but not in quasistatic flow, where $P(x_{\rm min})\!\sim\!x_{\rm min}^0$ was found. Our measurement of $P(x_{\rm min})$ shown in Fig.(\ref{xmin}) are strikingly similar to \cite{itamar}, and displays precisely these features, supporting the validity of our approach. Our finding that the relationship $\theta=1/\xi-1$ holds despite that the lack of strict independence  indicates that $\theta$ can be reliably extracted from size effects, and may thus be accessible experimentally.

{\it Conclusion:} The phenomenology of the depinning transition (see \cite{fisher, Ledoussal2002} for a review) of an elastic interface is very similar to that of the much less understood yield stress transition \cite{Zaiser2006,Talamali2011}: there is a critical force $F_c$ where the dynamics also consists of power-law avalanches. At larger forces, the velocity follows $V\sim (F-F_c)^\beta$, a form equivalent to the Herschel-Bulkley relation. There are several important differences however. First,  for depinning $\beta \le 1$  ($\beta=1$ is the mean field result), whereas  yield stress materials generally display $\beta>1$. Here we have shown that models where unstable sites can be re-stabilized, which occurs for non-monotonic interactions and $\tau_c>0$, can indeed present $\beta>1$. Second, in depinning the number of avalanches triggered when $F$ is slightly increased below $F_c$ is extensive and not singular near the transition. This fact allows one to obtain the exponent $\tau$ characterizing avalanches  \cite{Narayan1993, Aragon2012}. Near the yield stress transition, atomistic simulations \cite{salerno} support that this hypothesis does not hold, and that the number of avalanches triggered at $\sigma_c$ depends on the volume with a non-trivial exponent. We interpret that  difference as stemming from the fact that $\theta=0$ for depinning, whereas the long range and non-monotonicity of the interactions allow $\theta>0$ for the yielding transition, implying a non-trivial relationship between stress increase and number of avalanches triggered. This result supports that the exponent $\theta$ should enter in a yet-to be done scaling description of the yielding transition. 
This exponent also described materials after a isotropic quench, and may thus also play a role near the glass transition. This point of view  emphasizes the role of elastic interactions, often neglected in theories of the glass transition, but which have recently been proposed  to control the fragility of liquids \cite{Yan}.

Acknowledgments: it is a pleasure to  thank G.~D{\"u}ring, Le Yan, and Eric DeGiuli for comments on the manuscript. This work has been supported by
the Sloan Fellowship, NSF CBET-1236378, NSF DMR-1105387, and Petroleum Research Fund \#52031-DNI9. This work was also supported partially by the MRSEC
 Program of the National Science Foundation under Award Number DMR-0820341, and the hospitality of the Aspen Center for Physics.
\bibliography{reference7.bib}{}
\appendix{

\newpage
\section{Appendix: solution of the mean-field model
}
We give here a few more details about the solution of the mean-field case. We write the distribution $P(x)$ as $P(x)=P_1(x)+P_2(x)$, where $P_1(x)$ is the distribution of stable ($x>0$) and mechanically unstable ($x<0$) sites, and $P_2(x)$ is the distribution of collapsed sites. In the thermodynamic limit, $P_1$ and $P_2$ satisfy the Fokker-Planck equations 
\be
\frac{\partial P_1(x)}{\partial t}=\\
\dot\gamma\Big(D\frac{\partial^2P_1(x)}{\partial x^2}+\lambda \frac{\partial P_1(x)}{\partial x}+\delta(x-1)\Big)-\theta(-x)\frac{P_1(x)}{\tau_c}\label{AFP1}
\ee
\be
\frac{\partial P_2(x)}{\partial t}=\\
\dot\gamma\Big(D\frac{\partial^2P_1(x)}{\partial x^2}+\lambda \frac{\partial P_1(x)}{\partial x}\Big)+\theta(-x)\frac{P_1(x)}{\tau_c}-\frac{P_2(x)}{\tau_r}\label{AFP2}
\ee
In the stationary limit, these two equations imply, by taking the integral of ($\ref{AFP1}$) and ($\ref{AFP2})$, the following conservation law :
\begin{equation}
\label{conservation}
\dot{\gamma}=\frac{\int_{-\infty}^{\infty} P_2(x) dx}{\tau_r}=\frac{\int_{-\infty}^{0}P_1(x) dx}{\tau_c}
\end{equation}
This equality simply states that in a stationary state, the flux of collapsing site is equal to the flux of sites that become stabilized again. Solving ($\ref{AFP1}$) gives $P_1$ up to a constant, which we can then determine thanks to ($\ref{conservation}$), using the normalization of the complete distribution $P$. We find the critical value $\lambda_c=1$ above which there exists a stationary solution with $\dot{\gamma}>0$. The same equation $\ref{conservation}$ allows to solve for $\dot{\gamma}$ as a function of $\lambda-\lambda_c$. Expanding for $\lambda\simeq\lambda_c$ gives :
\be
\label{scaling1}
\dot{\gamma}=\frac{(\lambda-\lambda_c)^2}{D\tau_c}-\frac{(\tau_c+2\tau_r)(\lambda-\lambda_c)^3}{D^2\tau_c^2}+O(\lambda-\lambda_c)^4
\ee
We then have to relate $\lambda-\lambda_c$ and $\sigma-\sigma_c$, which is done using that $\int_{-\infty}^{+\infty} xP(x)dx=1-\sigma$. Multiplying ($\ref{AFP2}$) by $x$ and integration by parts yields :
\be
\displaystyle \int_{-\infty}^{+\infty} xP_2(x)dx=\frac{\tau_r}{D\tau_c}(\lambda_c-\lambda)^3
\ee
\noindent and because we have already computed $P_1$, we can compute the expansion of $\sigma-\sigma_c$ in terms of $\lambda-\lambda_c$.
A few straightforward computations finally yield ($\ref{scalingMF}$).

\end{document}